\documentclass[a4paper,10pt]{article}
\usepackage[dvips]{graphicx}
\usepackage{amsfonts,amsmath,amssymb}

\newcommand{\mc}[1]{\mathcal{#1}}
\newcommand{\f}[2]{\frac{#1}{#2}}

\begin{document}
%-------------------------------------------------------------------------------
\title{\bf Asymptotically Lifshitz Brane-World Black Holes}
\author{Arash Ranjbar\footnote{Electronic address: a\_ranjbar@sbu.ac.ir}, H. R. Sepangi\footnote{Electronic address: hr-sepangi@sbu.ac.ir}\quad and\quad S. Shahidi\footnote{Electronic
address: s\_shahidi@sbu.ac.ir} \\
\small Department of Physics, Shahid Beheshti University, Evin,
Tehran 19839 Iran}
\maketitle
\begin{abstract}
We study the gravity dual of a Lifshitz field theory in the context of a RSII brane-world scenario, taking into account the effects of the extra dimension through the contribution of the electric part of the Weyl tensor.
We study the thermodynamical behavior of such asymptotically Lifshitz black holes.
It is shown that the entropy imposes
the critical exponent $z$ to be bounded from above. This maximum value of $z$ corresponds to a positive infinite entropy as long as the temperature is kept positive. The stability and phase transition for different spatial topologies are also discussed.
\end{abstract}

%===============================================================================
%===============================================================================
\section{Introduction}\label{sec1}
%===============================================================================
%===============================================================================
Quantum field theories and their holographic duals in the context of
AdS/CFT have been the subject of numerous studies in recent years.
The AdS/CFT correspondence provides a well established method for obtaining
dual descriptions of strongly coupled conformal field theories in
terms of weakly coupled gravitational theories where the
gravitational background symmetries are realized as the symmetries
of the dual CFT. For example, the conformal group $SO(D,2)$ of a $D$
dimensional CFT arises as the isometry group of $AdS_{D+1}$.

The AdS/CFT seems to extend to some systems which do not exhibit
Lorentz invariance. Recently, attempts have been made to apply the
holographic principle to study condensed matter systems near a
critical point \cite{Sachdev,Hartnoll} where there are many strongly
coupled systems which have to be studied non-perturbatively. In this
situation, one encounters many scale-invariant field theories which
are not Lorentz invariant. In such theories, space and time are not
considered in the same footing, so they can scale in  different ways
under dilatation
\begin{align}\label{1}
t\rightarrow\lambda^zt,\quad \vec{x}\rightarrow\lambda\vec{x},
\end{align}
where $z$ is the so-called dynamical critical exponent showing
relative scale dimension of space and time. For $z=1$, this scaling
symmetry reduces to the familiar relativistic scale invariance
\begin{align}\label{eq2}
t\rightarrow\lambda t,\quad \vec{x}\rightarrow\lambda\vec{x}.
\end{align}
The gravitational dual of theories with non-relativistic scale invariant transformation (\ref{1}) has been
already considered with Schr\"{o}dinger symmetry \cite{Son,Bala} and Lifshitz symmetry  \cite{Kachru}.

In this paper, we focus our attention on the Lifshitz theory as an anisotropic
scale-invariant theory. A toy model which exhibits such anisotropic
scale invariance is the Lifshitz field theory represented by
\begin{align}\label{eq3}
{\cal {L}}=\int d^2xdt\bigg[\,(\partial_t\phi)^2-\kappa(\nabla^2\phi)^2\bigg].
\end{align}
The theory describes a line of fixed points parametrized by $\kappa$ and respects the
scaling invariance (\ref{1}) with $z=2$. Although, this anisotropic scaling theory is mostly applied to condensed matter theories, it has been shown recently that some models for particle physics could benefit of Lifshitz symmetry in their Lagrangians \cite{FTL}.

Generally speaking, theories  assuming scaling invariance as well as
time and space translations, spatial rotation, parity and time
reversal symmetries, admit a metric of the form \cite{Kachru,eff,coset}
\begin{align}\label{4}
ds^2=l^2\,\bigg(-r^{2z}dt^2+\frac{dr^2}{r^2}+r^2d\Omega_k^2\bigg),
\end{align}
where $l$ is a characteristic length scale.

The Einstein-Hilbert (EH) action cannot admit a
Lifshitz geometry \cite{Kachru,eff}.
However, there are generically three ways to construct actions which
admit such geometry (\ref{4}). One way is by adding some nontrivial
matter terms to the EH action \cite{Kachru} which can be shown to be equivalent to adding a massive gauge field \cite{balasub}.
It is generally difficult to obtain analytic solutions to the field
equations generated by these actions. Attempts have been made to
find analytic and numeric black hole/brane solutions which asymptotically admit
Lifshitz geometry \cite{Danielson,Mann,Peet,charge,3dimention,maeda,amado,super,brena}.
The embedding of such theories in the string theory and supergravity has been also discussed in \cite{string}.

As a second approach but not completely independent of the former,  one may
add higher order curvature terms to generate dynamics for parameter
$z$ to ensure that the Lifshitz geometry (\ref{4}) satisfies the
resulting field equations of the new action \cite{Cai,Pang,higher
dimension,Dehghani}.  We refer to these procedures as a renormalization
of $z$. Also recently it has been shown that stationary black hole solutions can be obtained in the context of this approach \cite{bayramjan}.

There is however another approach which introduces a dilaton field together with some $U(1)$ gauge fields instead of a massive gauge field \cite{marijoon}. This approach has an advantage respect to others, since finding analytic solutions are more probable. But from the holographic point of view, this approach has some difficulties in the sense that the dilaton field diverges on the boundary \cite{marijoon}.  Analytic solutions for asymptotically Lifshitz black holes/branes of this model have been obtained in \cite{utrekh}.

Also recently, a non-abelian  Lifshitz Chern-Simons gauge theory in $2+1$ dimensions with dynamical critical exponent $z=2$ has been proposed \cite{kachi} to explain the possibility of a phase transition from an isotropic to anisotropic fractional quantum hall systems. The gravitational dual of such theory was obtained in \cite{balabala}.

We propose an alternative to renormalize the dynamical exponent $z$,
considering the global effects of the extra dimensions.
Consequently, asymptotically Lifshitz black hole solutions can be obtained
in the context of brane-world scenarios which is the main purpose of
the present paper. For a review on the brane world scenarios see \cite{maart}.

In order to obtain
the equations of motion governing gravity on the brane one can use the method
initiated by Shiromizu, Maeda and Sasaki (SMS) \cite{sms}. They
assumed general bulk and brane geometries and obtained equations of
motion of the matter field on the brane by use of the Gauss-Codazzi
equations and Israel junction conditions \cite{israel}. The
resulting field equations are $4D$ and include the global bulk structure through the
electric part of the Weyl tensor. The main feature of these
equations is the introduction of a new tensor which is quadratic in
the energy-momentum tensor.

The brane generalization of the vacuum Einstein field equations (\ref{4})
does not admit the Lifshitz geometry.
Roughly speaking, this arises from the dependence of matter components
to the couplings of the theory. This implies that a Lifshitz space-time cannot
be admitted as a solution by the field equations on the brane in the absence of
an energy-momentum tensor. The calculations
lead to this result are presented in the Appendix. There, we show that the Lifshitz space-time can be embedded on the brane in the presence of matter field.
As we will show below, the brane field equations can admit
asymptotically Lifshitz geometry. This is done with the aid of the electric part
of the Weyl tensor which ensures the renormalizability of the dynamical
exponent $z$. The generalization of this work to the case that we have matter fields are straightforward. It is worth mentioning that attempts have been made to
show that the Lifshitz space-time can be a solution of
a brane-world model in the presence of an anisotropic perfect fluid
energy-momentum tensor on a thick brane \cite{Koroteev,Gordeli}.

The paper is organized as follows: In the next two sections we
obtain the asymptotically Lifshitz solution to the brane geometry
and show that the value of the dynamical exponent must be bounded
from below to represent a black hole solution. We
interpret the constants of the solution in terms of the mass and
charge of a black hole using the fact that in the case $z=1$ the
solutions reduce to the AdS-Reissner-Nordstr\"om black hole.  In section
\ref{sec4} we study the thermodynamical properties of such black
holes. As we will see, the entropy restricts the value of the
dynamical exponent from above. This result shows that in vacuum, the
asymptotically Lifshitz black holes in Randall-Sundrum brane scenario only admit a narrow range
for $z$. The stability of the black hole and the possibility of
phase transition in the most interesting case of $z=2$ will also be discussed.
The reasons for the lack of a vacuum Lifshitz solution in the present
framework are discussed in the Appendix. Conclusions are drawn in the last section.
%===============================================================================
%===============================================================================
\section{The setup}\label{sec2}
%===============================================================================
%===============================================================================
We begin with the field equations of the  $4D$ brane \cite{sms}
\begin{align}\label{eq5}
 G_{\mu\nu}=-\Lambda g_{\mu\nu}+\kappa_4^2
\tau_{\mu\nu}+\kappa_5^4\pi_{\mu\nu}-\mc{E}_{\mu\nu},
\end{align}
where
\begin{align}\label{eq5.1}
\pi_{\mu\nu}=-\frac 1 4
\tau_{\mu\alpha}\tau^\alpha_{~\nu}+\f{1}{12}\tau\tau_{\mu\nu}+\f{1}{8}g_{\mu\nu}
\tau_{\alpha\beta}\tau^{\alpha\beta} -\f{1}{24}g_{\mu\nu}\tau^2,
\end{align}
and $\tau_{\mu\nu} $ is the energy-momentum tensor of the matter
fields residing on the brane and $\mc{E}_{\mu\nu} $ is the electric
part of the Weyl tensor. The Greek indices in the above equations are defined on the brane and run
from 0 to 3. It is worth mentioning at this early stage that as has been mentioned in \cite{sms},  equation (\ref{eq5}) is not sufficient for obtaining the whole dynamics of the bulk. There are some
other equations from which one may obtain the bulk dynamics \cite{maart}. However, equation (\ref{eq5}) is sufficient to obtain the
dynamics of the brane, and the 5-dimensional equations are then used to obtain
the $y$-dependence of the bulk metric and the electric part of the Weyl tensor
where $y$ is the coordinate of the fifth dimension. In
the present paper, we shall consider the dynamics of the brane without worrying
about the $y$-dependence of the metric since this can be trivially found by a
Taylor series as in \cite{maart}.

Let us use the following irreducible decomposition for the electric
part of the Weyl tensor \cite{maart}
\begin{align}\label{eq9}
\mc{E}_{\mu\nu}=-\kappa^4\left[U\left(u_\mu
u_\nu+\f{1}{3}h_{\mu\nu}\right)+\mc{P}_{\mu\nu}+2\mc{Q}_{(\mu}u_{\nu)}\right],
\end{align}
where $\kappa=\f{\kappa_5}{\kappa_4}$, and the spatial metric
$h_{\mu\nu}=g_{\mu\nu}+u_\mu u_\nu$ is orthogonal to the chosen normalized
time-like vector field $u^\mu$ with the property $u^\mu u_\mu=-1$. In this decomposition, we introduce three
terms, known as  dark radiation
\begin{align}\label{eq10}
U=-\kappa^{-4}\mc{E}_{\mu\nu}u^\mu u^\nu,
\end{align}
dark energy flux
\begin{align}\label{eq11}
 \mc{Q}_\mu=\kappa^{-4}h^\alpha_\mu\mc{E}_{\alpha\beta}u^\beta,
\end{align}
and dark pressure
\begin{align}\label{eq12}
 \mc{P}_{\mu\nu}=-\kappa^{-4}\left[h_{(\mu}^\alpha
h_{\nu)}^\beta-\f{1}{3}h_{\mu\nu}h^{\alpha\beta}\right]\mc{E}_{\alpha\beta}.
\end{align}
In the case of a static and isotropic space-time we have $\mc{Q}_\mu=0$,  and
\begin{align}\label{eq13}
\mc{P}_{\mu\nu}=P(r)\left[r_\mu r_\nu-\f{1}{3}h_{\mu\nu}\right],
\end{align}
where $r_\mu$ is the unit radial vector \cite{maar1}.
%%%%%%%%%%%%%%%%%%%%%%%%%%%%%%%%%%%
\section{Asymptotically Lifshitz Black hole Solution}\label{sec3}
%%%%%%%%%%%%%%%%%%%%%%%%%%%%%%%%%%%
The asymptotically Lifshitz space-time can be described in general by
\begin{align}\label{eq6}
ds^2=-f(r)\left(\f{r}{l}\right)^{2z}dt^2+\f{1}{f(r)}\left(\f{l}{r}
\right)^2dr^2+r^2 d\Omega_k^2,
\end{align}
where $d\Omega_k^2$ is the $2D$ portion of the metric which depends on
the curvature index of the space-time
\begin{align}\label{eq7}
 d\Omega_k^2=d\theta^2+\left\{\begin{array}{ll}
                              \sin^2\theta d\phi^2\qquad& k=+1 \\
                  \theta^2d\phi^2\qquad& k=0 \\
                  \sinh^2\theta d\phi^2\qquad& k=-1
                             \end{array}
,\right.
\end{align}
and, as  was mentioned in the Introduction, $z$ shows the relative
scaling of the space and time. This form of the metric guarantees
that the brane geometry approaches  the Lifshitz geometry, equation (\ref{4}),
up to the re-scaling $r\rightarrow r/l$ and $t\rightarrow t/l$ if $f(r)\rightarrow 1$ as $r\rightarrow\infty$ .

We are interested in the vacuum solutions of the model, so we assume that the energy-momentum tensor is zero. As a result,
the brane equations of motion reduce to the form
\begin{align}\label{eq8}
G_{\mu\nu}+\Lambda g_{\mu\nu}+\mc{E}_{\mu\nu}=0.
\end{align}
Assuming $u^\mu=\delta^\mu_0$ and $r^\mu=\delta^\mu_1$, the electric
part of the Weyl tensor can be computed from  metric  (\ref{eq6})
as
\begin{align}\label{eq14}
\mc{E}^0_0&=\kappa^4 U(r),\qquad
\mc{E}^1_1=-\f{1}{3}\kappa^4\left[U(r)+2P(r)\right],\nonumber\\
 \mc{E}^2_2&=\mc{E}^3_3=-\f{1}{3}\kappa^4\left[U(r)-P(r)\right].
\end{align}
Using equations (\ref{eq6}), (\ref{eq8}) and (\ref{eq14}) one can
write the Einstein equations
\begin{align}\label{eq15}
&\f{rf^\prime+3f}{l^2}-\f{k}{r^2}+\kappa^4U(r)+\Lambda=0,\\
&\f{rf^\prime+(1+2z)f}{l^2}-\f{k}{r^2}-\f{1}{3}\kappa^4\bigg[U(r)+2P(r)\bigg]
+\Lambda=0,
\end{align}
and
\begin{align}
\f{3}{2}\f{1+z}{l^2}rf^\prime&+\f{z(1+z)}{l^2}f+\f{1}{2l^2}r^2f^{\prime\prime}\nonumber\\
&-\f{1}{3}\kappa^4\bigg[U(r)-P(r)\bigg]+\Lambda=0.
\end{align}
These equations can be solved for the metric coefficient with the result
\begin{align}\label{eq16}
 f(r)&=-\f{2\Lambda
l^2}{z^2+2z+3}+\f{k}{1-z+z^2}\left(\f{l}{r}\right)^2\nonumber\\
&+C_2\left(\f{l}{r}\right)^{B_2}+C_3\left(\f{l}{r}\right)^{B_3},
\end{align}
where we have defined
\begin{align}\label{eq16.1}
B_2&\equiv2+\f{3z}{2}+\sqrt{\f{z^2}{4}+2z-2}~,\nonumber\\
B_3&\equiv2+\f{3z}{2}-\sqrt{\f{z^2}{4}+2z-2}~,
\end{align}
with both $C_2$ and $C_3$ being constants. Solution (\ref{eq16}) will also satisfy the Bianchi identities as it should.

To have an asymptotic Lifshitz black hole, the
leading term of the function $f(r)$ must be equal to unity and the constants $B_2$ and $B_3$ must be well defined and positive. This means that in a brane-world scenario the dynamical exponent is
bounded from below,
\begin{align}\label{eq16.11}
z\geq-4+2\sqrt{6}.
\end{align}
Also, the cosmological constant could be read off as a function of the dynamical
exponent
\begin{align}\label{eq16.2}
\Lambda=-\f{1}{2}\f{z^2+2z+3}{l^2}.
\end{align}
The calculations would become much simpler if one writes $f(r)$ in the following form
\begin{align}\label{eq17}
 f(r)=\sum_{\mu=0}^{3}C_\mu\left(\f{l}{r}\right)^{B_\mu},
\end{align}
and defines
\begin{align}\label{eq18}
B_0=0,\quad B_1=2,\quad C_0=1,\quad C_1=\f{k}{1-z+z^2}.
\end{align}
It is now possible to calculate dark radiation and pressure from
equation (\ref{eq15}) with the result
\begin{align}\label{eq19}
\kappa^4
U(r)&=-\Lambda+\f{k}{r^2}-\sum_{\mu=0}^{3}\f{C_\mu(3-B_\mu)}{l^2}\left(\f{l}{r}
\right)^{B_\mu},\\
\kappa^4
P(r)&=2\Lambda-\f{2k}{r^2}+\sum_{\mu=0}^{3}\f{C_\mu(3-2B_\mu+3z)}{l^2}\left(\f{l}
{r}\right)^{B_\mu}.
\end{align}
Since the integration constants $C_2$ and $C_3$ are independent of $z$ they can be obtained by considering the
behavior of the solution at $z=1$. This is the case where the
relativistic scaling invariance is realized  and the resulting solution can be
compared with the known black hole solutions. In this case,
the asymptotic Lifshitz solution of the brane reduces to the form
\begin{align}\label{eq19.1}
ds^2=&-f_k(r)dt^2
+\f{1}{f_k(r)}dr^2+r^2
d\Omega^2,
\end{align}
with
\begin{align}
 f_k(r)=k+\f{r^2}{l^2}+\f{C_3l}{r}+\f{C_2l^2}{r^2},
\end{align}
where $d\Omega^2$ is given by equation (\ref{eq7}) and
$\Lambda=-\f{3}{l^2}$. Comparing with the standard topological
Reissner-Nordstr\"om-AdS black hole solution
\begin{align}\label{eq20}
ds^2=&-g_k(r)dt^2+\f{1}{g_k(r)}dr^2+r^2
d\Omega^2,
\end{align}
with
\begin{align}
g_k(r)=k-\f{\Lambda}{3}r^2-\f{2M}{r}+\f{Q^2}{r^2},
\end{align}
one can read off the constants as
\begin{align}\label{eq21}
C_2=\f{Q^2}{l^2}\equiv q^2,\qquad C_3=-\f{2M}{l}\equiv -2m.
\end{align}
From now on, $m$ and $q$ are referred to as the mass and charge parameters
of the general asymptotic Lifshitz black hole \eqref{eq19.1}. It is worth to mention that the charge here does not mean a real charge because the theory does not contain the electromagnetic source. However, one can expect that the electric part of the Weyl tensor is responsible for the appearance of such term.

We also mention that the position of the horizon $r_H$ of the
black hole can be obtained through imposing the condition $f(r_H)=0$,
which for the solution \eqref{eq17} leads to
\begin{align}\label{eq21.1}
u^{2+\f{3z}{2}}+C_1 u^{\f{3z}{2}}+q^2 u^{-b} -2m u^b=0,
\end{align}
where we have defined
\begin{align}\label{eq21.2}
b=\sqrt{\f{z^2}{4}+2z-2},\qquad u=\f{r_H}{l}.
\end{align}
One can easily verify numerically, that the above equation has solutions for all allowed values of $z$. Hereafter, this premise is eligible for future use in thermodynamics.
%%%%%%%%%%%%%%%%%%%%%%%%%%%%%%%%%%%%%%%
\section{Thermodynamical properties}\label{sec4}
%%%%%%%%%%%%%%%%%%%%%%%%%%%%%%%%%%%%%%%%
In this section we study the thermodynamical behavior of the
asymptotically Lifshitz black holes (\ref{eq6}). After calculating
the corresponding thermodynamical quantities for such black holes we
discuss their stability.

The gravitational mass of a black hole, $M$, can be obtained by the
equation $f(r_H)=0$, governing the position of the horizon. In the
case of black hole (\ref{eq6}) we find
\begin{align}\label{eq22}
M=\f{l}{2}u^{B_3}\sum_{\mu=0}^2 C_\mu u^{-B_\mu},
\end{align}
where $u$ is associated with the outer most horizon.

The Hawking temperature associated with a black hole is obtained by
requiring the absence of a conical singularity at the black hole
horizon in the Euclidean sector of the asymptotically Lifshitz black
hole \cite{frolov}. One then has
\begin{align}\label{eq25}
T=\f{\kappa}{2\pi},
\end{align}
where the surface gravity $\kappa$ on the horizon admits the following useful
representation
\begin{align}\label{eq24.1}
 \kappa^2=-\f{1}{2}\xi_{\alpha;\beta}\xi^{\alpha;\beta},
\end{align}
where $\bf{\xi}$ is a killing vector normal to the surface of the
horizon. As a consequence, the surface gravity of a black hole could
be written in terms of the temporal component of the metric as
\begin{align}\label{eq24}
\kappa=-\f{1}{2}\f{\textrm{d}g_{tt}}{\textrm{d}r}\bigg\lvert_{r=r_H}.
\end{align}
The Hawking temperature of the asymptotically Lifshitz black hole
(\ref{eq6}) is now given by
\begin{align}\label{eq25.1}
T=\f{1}{4\pi l}u^{2z-1}\sum_{\mu=0}^2 C_\mu(B_3-B_\mu)u^{-B_\mu}.
\end{align}
One should note that for asymptotically flat space-times, the normalization of
the killing vector field, $\bf{\xi}$, in (\ref{eq24.1}) is imposed at the spatial infinity.
However, it has been shown by Brown \textit{et al.} that the choice of the
normalization elsewhere results in the appearance of an additional redshift factor,
the so-called Tolman redshift factor, for a given temperature in a stationary
gravitational field \cite{brown}. For non-asymptotically flat space-times, one
should impose the normalization at a finite space-like boundary surface. Nevertheless,
one can assume that observations take place at the boundary surface where
the redshift factor equals unity. So, in what follows we set the Tolman redshift factor
equal to ``one'' without loss of generality.

Another important thermodynamical quantity is the entropy of black
holes. In Einstein gravity the entropy of a black hole is obtained
by the so-called Bekenestein area formula. This means that the entropy is equal
to one-quarter of the horizon area. It has also  been shown that
when higher curvature terms are present, this statement is no longer
true \cite{high,noji}. However, Wald \cite{wald} has shown that in any
case the entropy of a black hole is a function of the horizon
geometry.

Alternatively, given the temperature of a black hole as a function of
energy (or as a function of mass $E=M$), we can integrate the first law of thermodynamics, $dE=TdS$,
to which a black hole as a thermodynamical system should conform,
to define an ``entropy'' for the black hole
\begin{align}
S&=\int_0^{M}\f{1}{T}\textrm{d}E.
\end{align}
It is suitable to write it as an expression which is useful for future application, so
 we change the integration variable to radial coordinate
\begin{align}\label{eq26}
S=\int_0^{r_H}\f{1}{T}\f{\partial M}{\partial r}\textrm{d}r,
\end{align}
where the upper limit of integration is the horizon radius.

It is worth to mention that the term containing the charge does not appear in the first law of thermodynamics, because we assume that the spurious charge introduced in the theory is kept constant among the thermodynamical analysis.

Using (\ref{eq22}), (\ref{eq25.1}) and (\ref{eq26}) we find
\begin{align}\label{eq26.01}
S=\f{2\pi l^2}{B_3-2z+1}u^{B_3-2z+1}.
\end{align}
The positivity condition on the entropy implies an upper bound on
the dynamical critical exponent $z$, that is
\begin{align}\label{eq26.1}
z< \f{11}{5}.
\end{align}
From equation (\ref{eq26.01}) we see that  $z=\f{11}{5}$ corresponds
to an infinitely large entropy. It then implies that the dynamical
exponent $z$ must lie near this value by the maximum entropy
principle. This behavior however resembles some condensed matter systems such as
Rokhsar-Kivelson dimer model \cite{rokh} and the strongly correlated electron systems \cite{strong}, which is well described
by the action \eqref{eq3} at $z=2$. From the view point of duality, our result is expectable since our gravitational model
becomes strongly stable near the point for which its dual theory becomes stable and physical. It is worth mentioning that the limit \eqref{eq26.1} is obtained in the context of asymptotically Lifshitz space-times which may induce a perturbation term on the dual field theory \eqref{eq3}. This perturbation term can cause the critical exponent admitting values up to $z=\f{11}{5}$, supporting our gravitational model. As a result, our model restricts the physically reasonable
solution to those which admit a dynamical exponent $z$ in the
range
\begin{align}\label{eq26.2}
-4+2\sqrt{6} \leq z < \f{11}{5}.
\end{align}
However, if we do not take the condition (\ref{eq26.1}) \textit{ab
initio}, the entropy will contain negative values. This happens, for
example, in the case of the Gauss-Bonnet black holes for negative
spatial curvature. It has been emphasized
that negative values for the entropy is related to the lower limit of
the integral (\ref{eq26}). This problem could be resolved by considering
a minimum radius $r_{min}$ corresponding to the zero gravitational mass
as the lower limit of the integral \cite{high,odi}. In the case of the black hole
(\ref{eq6}), the entropy is independent of the spatial curvature so
that it can be negative for all topologies. In our case instead of introducing the lower limit of integral we
remove negative values for the entropy by restricting the dynamical
exponent $z$.

In order to study the thermodynamical stability of black holes, one
should calculate the heat capacity. The local thermal stability of a
black hole is determined by the sign of the heat capacity. If the
heat capacity is positive, the black hole is locally stable to
thermal fluctuations. Negative values for the heat capacity result
in unstable black holes. We
define the heat capacity for a constant charge black hole as
\begin{align}\label{eq27}
C_Q\equiv\left(\f{\partial M}{\partial T}\right)_Q=\left(\f{\partial M}{\partial
r_H}\right)_Q\left(\f{\partial r_H}{\partial T}\right)_Q.
\end{align}
For the black hole (\ref{eq6}), the heat capacity in constant charge can be written as
\begin{align}\label{eq28}
C_Q=&2\pi l^2u^{B_3-2z+1}\f{\sum_{\mu=0}^2 C_\mu(B_3-B\mu)u^{-B_\mu}}
{\sum_{\mu=0}^2 C_\mu(B_3-B\mu)(2z-1-B_\mu)u^{-B_\mu}}.
\end{align}
This expression may be converted to a more familiar thermodynamical relation
\begin{align}\label{eq29}
\f{1}{C_Q}=\f{1}{2\pi l^2}u^{2z-B_3}\left[\f{\partial}{\partial u}\ln(4\pi
T)\right].
\end{align}
The free energy of a black hole is defined as $F=M-TS$.
Using equations (\ref{eq22}), (\ref{eq25.1}) and (\ref{eq26}) which are
the expressions for gravitational mass, temperature and entropy
respectively, one obtains the free energy
\begin{align}
F=\f{l}{2}u^{B_3}\sum_{\mu=0}^2
C_\mu\left[\f{B_\mu-2z+1}{B_3-2z+1}\right]u^{-B_\mu}.
\end{align}
The value of the free energy determines the strength of the stability of
a black hole; a more stable black hole corresponds to less free
energy.

Moreover, the
singularity in the heat capacity corresponds to a phase transition
provided that the temperature and free energy possess a local
minimum and a local maximum at the specified points respectively.

Since the Lifshitz action \eqref{eq3} is scaling invariant for $z=2$, we are interested in focusing on this case.  Many condensed matter systems which contain quantum critical phenomena such as strongly correlated electron systems are described by the Lifshitz action \eqref{eq3} in $z=2$.
In addition, from the thermodynamical point of view, the case $z=2$  mimics the behavior of
black holes with maximum entropy.

In this case the normalized
temperature, the free energy and the heat capacity reduce to
\begin{align}\label{eq}
\Theta&=-\f{u^3}{12\pi}\left(3\sqrt{3}-15-\f{k}{u^2}\left(3-\sqrt{3}\right)+\f{
6q^2\sqrt{3} }{u^{3+\sqrt{3}}}\right),\nonumber\\
f&=\f{9u^{5-\sqrt{3}}+u^{3-\sqrt{3}}k-3q^2\left(2+\sqrt{3}\right)u^{-2\sqrt{3}}}
{6(-2+\sqrt {3})},\nonumber\\
c_q&=\f{2\pi}{
u^{\sqrt{3}}}\f{\left(3\sqrt{3}-15\right)u^{2}-k\left(3-\sqrt{3}\right)+6\sqrt{3
}q^2u^{-3-\sqrt{3}}}{
9\sqrt{3}-45-ku^{-2}\left(3-\sqrt{3}\right)-6q^2\left(2\sqrt{3}+3\right)},
\end{align}
where we define
\begin{align}\label{eq}
\Theta=Tl,\quad c_q=\f{C_Q}{l^2},\quad f=\f{F}{l}.
\end{align}
The thermodynamical behavior of the black hole differs
considerably depending on the charge of the black hole. Black holes
with a small amount of charge ( $q\leq 0.001$) show a phase transition, unlike
black holes with a large amount of charge which do not.

\begin{figure}
\centering
\includegraphics[scale=1]{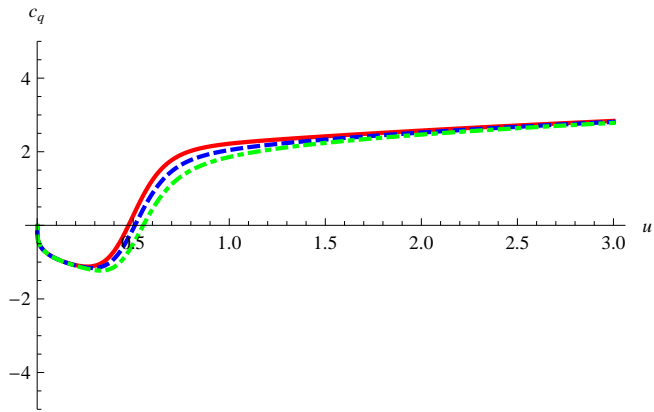}
\caption{Normalized heat capacity $c_q$ for $z=2$ and $q=0.1$ versus
the event horizon of the black hole. The solid, dashed and
dot-dashed curves correspond to $k=1$, $k=0$ and $k=-1$
respectively.}\label{HC in z equals 2}
\end{figure}

\begin{figure}
\centering
\includegraphics[scale=0.9]{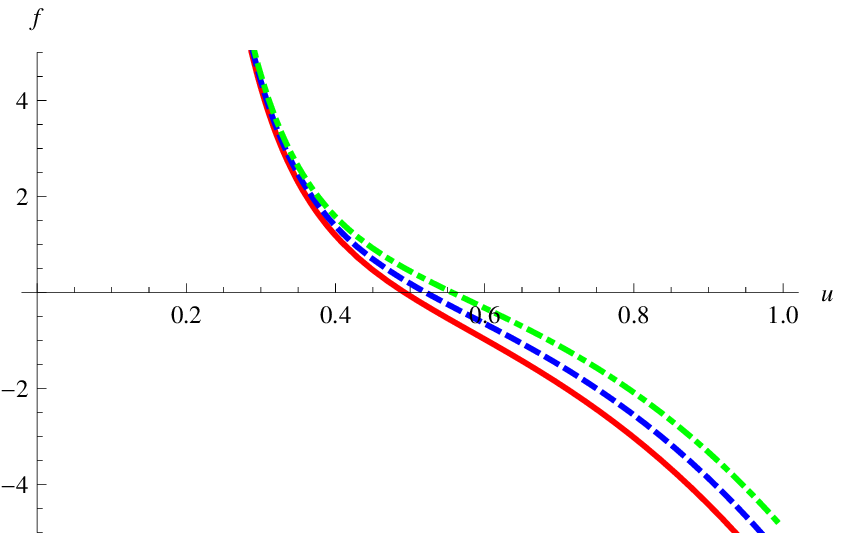}
\caption{Normalized free energy $f$ for $z=2$ and $q=0.1$ versus the
event horizon of a black hole. The solid, dashed and dot-dashed
curves correspond to $k=1$, $k=0$ and $k=-1$
respectively.}\label{free energy in z-2}
\end{figure}

We first consider the case $q=0.1$. The general behavior of black
holes with a large amount of charge are very similar to this case. Figure
\eqref{HC in z equals 2} shows the behavior of the heat capacity for
different spatial curvatures in this case. One can see that for all
horizon topologies with $k=\pm1,0$, black holes are stable beyond $u\sim
0.5$. As mentioned above, this value for the charge of a black hole
does not imply a phase transition which can be inferred  from the
smoothness of the heat capacity function.
Figure \eqref{free energy in z-2} shows the normalized free energy
for all spatial topologies. One can see from the figure that a
larger black hole is more stable.
\begin{figure}
\centering
\includegraphics[scale=0.9]{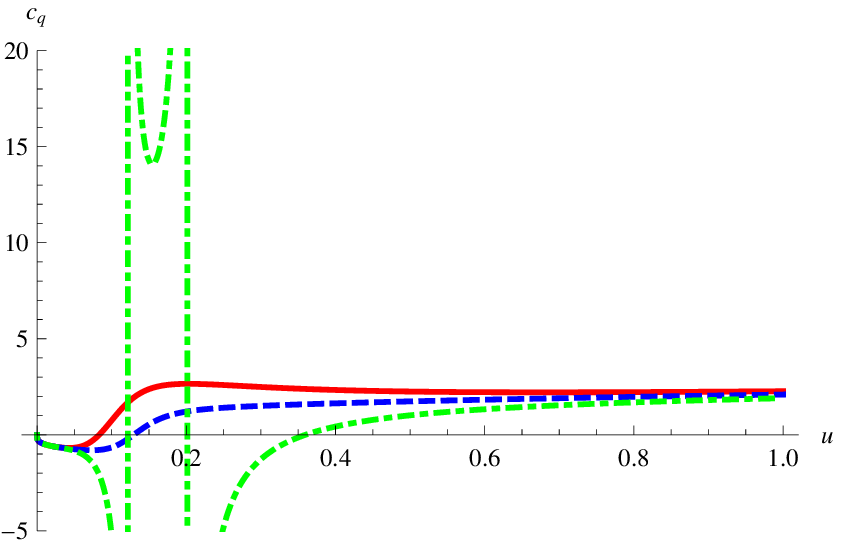}
\caption{Normalized heat capacity $c_q$ for $z=2$ and $q=0.001$
versus the event horizon of a black hole. The solid, dashed and
dot-dashed curves correspond to $k=1$, $k=0$ and $k=-1$
respectively.}\label{HC in z equals 2 and Q equals 0.001}
\end{figure}
\begin{figure}
\centering
\includegraphics[scale=1]{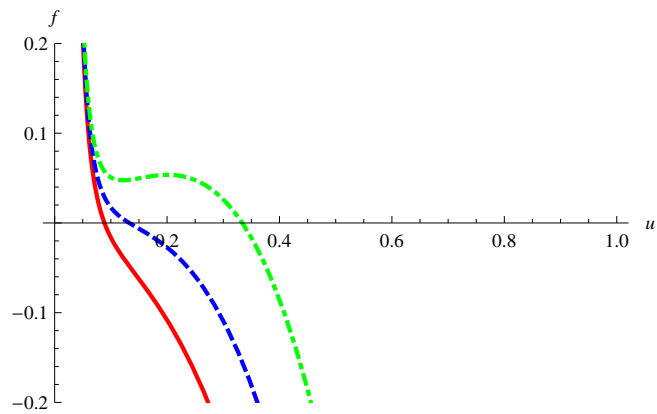}
\caption{Normalized free energy $f$ for $z=2$ and $q=0.001$ versus
the event horizon of a black hole. The solid, dashed and dot-dashed
curves correspond to $k=1$, $k=0$ and $k=-1$
respectively.}\label{free energy in z equals 2 and Q equals 0.001}
\end{figure}
The most interesting case is when the
normalized charge is less than $q\sim0.001$. Among such black
holes the most interesting ones are topological black holes
($k=-1$) where a phase transition could take place in principle.

It is appropriate to remark on the conditions for the occurrence of any physical phase transitions in Lifshitz theories.
It has been shown in \cite{super} that in order to have a physical phase transition in Lifshitz theories which is described by \cite{Kachru}, one must have
another length scale in addition to the Lifshitz length scale. This additional length scale was achieved  by introducing
new Maxwell fields into the theory. Nonetheless, the brane-world scenario discussed here has an intrinsic length scale introduced
by the ratio of the brane to the bulk Planck scales. In this context therefore, it is possible to have a physical phase transition.

For such black holes, there are two singularities in heat capacity.
However the regions between these two singular points
have negative temperature. The issue of a negative temperature for
black holes has been  investigated, for example, in \cite{PT in HL, Davis}
where the physical meaning of the phase transition for such black
holes is also discussed. The behavior of this type of a black hole is
shown in figure \eqref{HC in z equals 2 and Q equals 0.001}, for the normalized heat capacity,  and figure \eqref{free
energy in z equals 2 and Q equals 0.001}, for the normalized free energy . In cases $k=0$ and $k=1$,
black holes with a horizon larger than the approximate value
$u=0.1$ are stable so that the larger they get the more
stable they become. For the case $k=-1$, black holes with a horizon
radius in the outer region are more stable.

The maximum possible value for $z$, $\left(z=\f{11}{5}\right)$ corresponds to the infinitely
positive entropy and this indicates that the critical exponent must
eventually go close to the maximum value because of the maximum entropy
principle. In this case the free energy is minus infinity everywhere
as long as the temperature is positive. This suggests that in this case black holes are stable for positive
temperatures, independent of the horizon radii.

%=============================================================
%=============================================================
\section{Conclusions}
%=============================================================
%=============================================================
The Lifshitz space-time is
inconsistent with standard general relativity. Attempts have been
made to generalize GR in such a way as to admit  Lifshitz
space-times. In this paper we  have introduced another
generalization, i.e. higher dimensions, to deal with this problem.
Alas, the Lifshitz space-time is not the solution of the vacuum
brane field equations. However Lifshitz space-time can be admitted by the brane-world scenario with some ordinary matter fields as we will show in the appendix. We have shown that one can obtain an
asymptotically Lifshitz solution in a brane-world scenario with the
aid of the electric part of the Weyl tensor. The electric part of
the Weyl tensor  reflects the global geometry of the bulk
in brane-world scenarios whose effect is to enable the brane
admitting an asymptotically Lifshitz geometry. The electric part of the Weyl tensor on the other hand, can be realized as a probable source of some spurious charge. This tensor is however geometrical, so one can expect that the charge is not physical.  Therefore, one can assume that this charge is fixed in the thermodynamical considerations. From the
thermodynamical considerations we found that the dynamical exponent
must be bounded from above. Together with the lower limit for the
dynamical exponent $z$ which is implied by the existence of the
metric, we found that the dynamical exponent of the asymptotically
Lifshitz space-time cannot possess any arbitrary values. As a
result, the maximum value for $z$, that is, $z=\f{11}{5}$ corresponds to
the case where the entropy becomes infinite. From the maximum
entropy principle, we then expected that an asymptotic Lifshitz
black hole should approach the maximum value of $z$ where the
entropy is maximum. The case $z=2$ was used to find the
behavior of black holes in this case. In this case, for $q\lessapprox 0.001$ the black hole encounters a phase transition. It is worth to mention that the occurrence of phase transition in asymptotically Lifshitz space-time lies on the existence of another length scale beyond the characteristic Lifshitz length scale \cite{super}. In our model, it can be understood by the existence of a new length scale provided by the ratio of the bulk to the brane Planck scales.  The case $z=2$ also has another
interesting property for $k=-1$ in that the region between the two
singularities in heat capacity has a negative temperature. The
physical meaning of these cases are not clear yet.
%=============================================================
\section{Appendix}
%=============================================================
The standard $4D$ Einstein field equations do not admit the Lifshitz space-time
even in the presence of a perfect fluid. We first note that, as we mentioned in the Introduction,
the Lifshitz solution is obtained in thick brane models with an anisotropic perfect fluid.
In this Appendix we show that the RS brane-world scenario can admit the Lifshitz space-time as a
solution, in the presence of a perfect fluid.

The energy-momentum tensor of a perfect fluid is defined as
\begin{align}\label{eq-a-1}
\tau_{\mu\nu}=\rho(r) u_\mu u_\nu+p(r) h_{\mu\nu},
\end{align}
where $h_{\mu\nu}=g_{\mu\nu}+u_\mu u_\nu$. Using equation (\ref{eq5.1}), the tensor $\pi_{\mu\nu}$ reduces to
\begin{align}\label{eq-a-2}
\pi_{\mu\nu}=\f{1}{12}\rho\big[\rho u_\mu u_\nu+\left(\rho+2p\right)h_{\mu\nu}\big].
\end{align}
Also, the electric part of the Weyl tensor is given by equation (\ref{eq14}).
The resulting field equations are obtained using equation (\ref{eq5}) with the result
\begin{align}\label{eq-a-3}
&\f{3}{l^2}+\Lambda+\rho\left(\kappa_4^2+\f{\kappa_5^4}{12}\rho\right)+\kappa^4U=0,\nonumber\\
&\f{2z+1}{l^2}+\Lambda-\kappa_4^2 p-\f{\kappa_5^4}{12}\rho\left(\rho+2p\right)-\f{\kappa^4}{3}\left(U+2P\right)=0,\nonumber\\
&\f{z^2+z+1}{l^2}+\Lambda-\kappa_4^2 p-\f{\kappa_5^4}{12}\rho\left(\rho+2p\right)-\f{\kappa^4}{3}\left(U-P\right)=0.
\end{align}
The solution for the above system is
\begin{align}\label{eq-a-4}
p&=-\rho=6\f{\kappa^2}{\kappa_5^2},\quad\Lambda=-\f{z^2+2z+3}{l^2}+6\kappa^4,\nonumber\\
U&=\f{z^2+2z-3}{2\kappa^4l^2},\quad P=\f{z(1-z)}{\kappa^4l^2}.
\end{align}
This result implies that a RS brane, in the presence of a perfect fluid, admits the Lifshitz space-time.
However, the Lifshitz space-time is not a solution to the vacuum field equations. This becomes more
clear if one notices that the equation related to the energy density and pressure of the perfect fluid
depends on the coupling constants of the theory, which is exactly the reason why the Lifshitz space-time
cannot be a vacuum solution in brane-world scenarios.

%=============================================================
%=============================================================

\end{document}